\documentclass[11pt]{article}
\usepackage[T1]{fontenc}
\usepackage{lmodern}
\usepackage[margin=1in]{geometry}
\usepackage{amsmath,amssymb,amsthm,mathtools}
\usepackage{microtype,booktabs,array,enumitem}
\usepackage{xcolor}
\usepackage[colorlinks=true,linkcolor=blue!45!black,citecolor=blue!45!black,urlcolor=blue!45!black]{hyperref}
\hypersetup{pdftitle={Improvements in online transport},pdfsubject={Deterministic online transportation and shifted-scale analysis}}
\newtheorem{theorem}{Theorem}
\newtheorem{lemma}[theorem]{Lemma}
\theoremstyle{remark}

\newcommand{\OPT}{\operatorname{OPT}}
\newcommand{\ALG}{\operatorname{ALG}}
\newcommand{\down}{\operatorname{down}}
\newcommand{\fracpart}{\operatorname{frac}}

\newcommand{\Z}{\mathbb{Z}}

\setlist{itemsep=3pt,topsep=5pt}
\title{Online Transportation via Shifted Geometric Scales}
\author{John Jeang}
\date{}

\begin{document}
\maketitle
\vspace{-2.5em}
\begin{abstract}
In the Online Transportation problem, introduced by Kalyanasundaram and Pruhs in \cite{kalyanasundaram1995online}, requests arrive one at a time in a metric space and must immediately be assigned to one of $m$ facilities with unused capacity. Assignments are irrevocable, and the objective is to minimize the total distance traveled between request and facilities. 

The Online Transportation problem is a generalization of the Online Metric Matching problem, in which facilities have unit capacity, and inherits a $2m-1$ lower bound for all deterministic algorithms \cite{kp93,kmv94}. In \cite{kp93, kmv94}, it is shown that the Retrospective algorithm achieves a matching upper bound of $2m-1$ on the competitive ratio for the Online Metric Matching problem, and it is conjectured that there exists an algorithm achieving the same competitive ratio for the more general Online Transportation problem \cite{kp98}. The most recent progress on the problem is due to Seth, Das, and Raghvendra, who achieved a competitive ratio of $\approx6.66m - 2.89$ using their Robustified Greedy algorithm \cite{sdr26}. In this work, we give an algorithm with competitive ratio $\rho(m-1)+1$, where $\rho\approx5.64$. The algorithm builds on priority-based virtual DFS searches, based on a minimum spanning tree of the known facilities, similar to those used in \cite{hi25, ampu26}. However our algorithm uses a \textit{phase} to shift the geometric search scales, to decide when to skip certain searches and optimize the thresholds that define what edges to include in a given search.
\end{abstract}

\section{Introduction}
\label{sec:introduction}

Online transportation captures a basic allocation decision: a request needs service now, but assigning it consumes a resource that a future request may need more urgently. For example, a parking service must direct each arriving driver to a garage with an available space. Assigning the driver to the nearest available garage minimizes the immediate travel cost, but may exhaust capacity near a later driver's destination. Similar decisions arise when assigning demand to inventory at fixed depots or allocating a fixed batch of service slots among geographically distributed facilities. The model treats each assignment as consuming one unit of capacity for the duration of the input; it does not include departures or replenishment.

Formally, let $(X,d)$ be a metric space, where $X$ is the set of possible locations and $d$ is the distance function. There are $m$ distinct facilities, which we call garages, at the locations $G=\{v_1,\ldots,v_m\}\subseteq X$. Garage $v_i$ has positive integer capacity $c_i$. Write $c(v_i)=c_i$ and $C_{\mathrm{tot}}=\sum_{i=1}^m c_i$ for the total capacity. A sequence of $N\le C_{\mathrm{tot}}$ requests arrives at locations $x_1,\ldots,x_N$. On seeing $x_t$, an online algorithm must assign it immediately and irrevocably to a garage with unused capacity. Assignment to $h\in G$ costs $d(x_t,h)$. The garage locations, their capacities, and their mutual distances are known before the first request; future request locations are not known.

Let $\ALG$ denote the total cost of an online algorithm and $\OPT$ the minimum total cost of a feasible assignment that knows the entire request sequence. A competitive ratio of $R$ means that $\ALG\le R\OPT$ for every input. The goal is a guarantee in terms of $m$, independent of $N$ and the capacities. This distinction matters: replacing each garage by separate unit-capacity copies produces $C_{\mathrm{tot}}$ servers, which can be much larger than the number of distinct locations.

\subsection{Progress on the problem}

When every garage has unit capacity, the problem is Online Metric Matching. Classical work of Kalyanasundaram and Pruhs~\cite{kp93} and Khuller, Mitchell, and Vazirani~\cite{kmv94} establishes the optimal deterministic competitive ratio $2m-1$ for this special case. The same lower bound applies to transportation with arbitrary capacities because unit capacities are allowed. Kalyanasundaram and Pruhs conjectured that the upper bound $2m-1$ should also be achievable for transportation~\cite{kp98}. The conjecture asks whether having many capacity units at a few locations can be handled without increasing the optimal deterministic ratio.

Harada and Itoh~\cite{hi25} obtained the first deterministic guarantee linear in the number of locations: their Subtree-Decomposition algorithm is $(8m-5)$-competitive. Stephen Arndt, Benjamin Moseley, Kirk Pruhs, and Marc Uetz~\cite{ampu26} subsequently gave an $(8m-7)$-competitive algorithm in \emph{Competitive Online Transportation Simplified}, published at SOSA 2026. Their framework maintains virtual cars that may displace one another while preserving the garage occupancies of an irrevocable online assignment. This separates the analysis of a flexible virtual process from the implementation of actual decisions.

The 2026 preprint of Ritesh Seth, Syamantak Das, and Sharath Raghvendra~\cite{sdr26}, \emph{A Robustified Greedy Algorithm for Online Transportation with Improved Competitive Guarantees}, reports a deterministic ratio of approximately $6.66m-2.89$ (with the coefficients rounded) for its Robustified Greedy algorithm. It develops the Robust Matching approach and also reports a guarantee that depends on the geometry in fixed-dimensional Euclidean spaces. These additional guarantees concern features beyond the general-metric coefficient considered here.

The algorithm analyzed in this paper retains the virtual-search framework of Arndt et al.\ but changes its scale selection and initial priorities. We prove the ratio $\rho(m-1)+1$, where $\rho\approx5.64$ is defined exactly in Section~\ref{sec:constants}. The improvement concerns the leading coefficient in the general-metric bound.

\begin{center}
\small
\begin{tabular}{@{}p{0.63\textwidth}l@{}}
\toprule
Deterministic algorithm or analysis & Competitive guarantee \\
\midrule
Harada--Itoh (ICALP 2025)~\cite{hi25} & $8m-5$ \\
Arndt--Moseley--Pruhs--Uetz (SOSA 2026)~\cite{ampu26} & $8m-7$ \\
Seth--Das--Raghvendra (2026 preprint)~\cite{sdr26} & $\approx6.66m-2.89$ \\
This analysis & $\rho(m-1)+1$ \\
\bottomrule
\end{tabular}
\end{center}
Decimal values throughout this paper are rounded approximations. The proofs use exact expressions for all constants.

Section~\ref{sec:overview} explains the strategy and virtual exchanges. Section~\ref{sec:description} specifies the algorithm, and Section~\ref{sec:analysis} proves its competitive guarantee.

\subsection{Related work}

Robust Matching provides another approach to online metric matching. Raghvendra~\cite{r16} introduced a primal-dual algorithm with guarantees in both adversarial and random-arrival models. Nayyar and Raghvendra~\cite{nr17} developed an input-sensitive analysis that exploits the geometry of the server locations. The Robustified Greedy preprint~\cite{sdr26} continues this direction for capacitated transportation. Our proof instead uses bottleneck components of a spanning tree and threshold counts for virtual priorities.

Resource augmentation offers a complementary way to study transportation: the online algorithm is allowed more capacity than the offline comparator. Arndt, Ascher, and Pruhs~\cite{aap23} show that nearest-available greedy is $(1+2/(\lambda-2))$-competitive when its capacities are multiplied by a factor $\lambda\ge3$. Here the online algorithm and the comparator have identical capacities.

\section{Proof Overview}
\label{sec:overview}

The algorithm organizes a search for capacity around a spanning tree of the garage locations. The main idea is to let simulated cars exchange places while committing each real request only once. As formally stated, in the Online Transportation problem, a car is irrevocably committed to a garage upon arrival, however in the next section we justify why these exchanges are legitimate, then introduce the search terminology and describe the strategy.

\subsection{Why virtual evictions are valid: a swapping argument}
\label{sec:swapping-overview}

The algorithm maintains two configurations. The \emph{actual configuration} records the irrevocable assignments already made. The \emph{virtual configuration} is a simulation in which cars may be evicted and continue searching. After each arrival, the two configurations have the same number of occupied spaces at every garage, although the identities of the cars in those spaces may differ.

For a simple example, suppose an earlier car occupies a garage $g$ and a new request arrives at a location $x$. In the simulation, the new virtual car reaches $g$, takes the earlier virtual car's place, and sends that car searching until it finds a free space at another garage $h$. To implement this exchange, leave the earlier real car at $g$ and assign the new real request directly to $h$. Both configurations have the same occupancies: the occupancy at $g$ is unchanged, and one additional space is used at $h$. The swap concerns which virtual car represents the assignment, not an alteration to an earlier real decision.

The same argument applies to a chain of evictions. Each replacement preserves the occupancy of its garage and passes the search for an additional space to the displaced virtual car. Only the final parking uses a new space. The real newcomer is assigned to that final garage, which also has a free space in the actual configuration.

This construction also controls the cost. Concatenate the new virtual car's initial movement with the subsequent movements along the eviction chain. The result is a continuous walk from the new request to its actual destination: every handoff occurs at the garage where the previous virtual car stopped. By the triangle inequality, the direct actual assignment costs no more than the length of this walk. Thus virtual travel can pay for the real assignments even though the virtual identities change. Lemma~\ref{lem:implementation} makes these occupancy and cost arguments formal.

\subsection{Terminology: origins, priorities, and tours}
\label{sec:terms-overview}

\begin{description}[style=nextline,leftmargin=0pt,labelwidth=0pt,labelsep=0pt]
 \interlinepenalty=10000
 \item[Origin.] Each virtual car has a fixed origin: the nearest \textit{garage} to its request location, regardless of whether that garage is full. We emphasis that here ``origin" does \textbf{not} correspond to the location in the metric space in which a car ``begins". All of that car's searches are organized around this origin.
 \item[Search scale and phase.] A scale is a positive threshold drawn from a sequence that doubles at every step. The phase shifts the whole sequence multiplicatively. Tree edges are rounded downward to these scales; a search at a given scale may use only edges whose rounded lengths do not exceed that scale.
 \item[Search component.] Restricting to a subset of tree edges that fits a certain scale, on a given tree, gives a forest. A car searches the connected component containing its origin. Increasing the scale may join components and allow additional garages to be searched. This component need not be a metric ball around the origin.
 \item[Priority.] A car's priority is its current search scale, with an initial value of zero when the request is at its origin. It determines both the component being searched and the outcome of a conflict at a full garage. A smaller numerical priority wins against a larger one. The value never decreases for an individual car.
 \item[Tour.] A tour is a fixed depth-first walk through the search component, starting and ending at the origin and traversing each component edge twice. It tests garages for parking opportunities along the way. A parked car saves its position; if later evicted, it resumes rather than repeating the completed part of its tour.
 \item[Active and parked cars.] Exactly one virtual car is searching at a time while an arrival is processed. This is the active car. An eviction parks the active car and makes the displaced car active. Parked cars retain their priorities and saved tour positions.
\end{description}

\subsection{Algorithm overview and the main proof ideas}
\label{sec:strategy-overview}

Before any requests arrive, construct a minimum spanning tree using the original distances. Choose a phase for the doubling scales and a multiplier that converts the distance to a nearest garage into an initial search priority. These choices depend only on the known garage locations. Rounded edge lengths govern access to components; travel is always charged using original edge lengths.

On each arrival, the strategy is as follows.
\begin{enumerate}[label=\arabic*.]
 \item \textbf{Start near the request.} Introduce a virtual car at a nearest garage. Its initial priority is the rounded value of the multiplier times the distance from the request to that garage.
 \item \textbf{Search the current component.} Follow the tour at the current priority. Park at a free space when one is found. At a full garage, a car may replace a resident of largest priority if its own priority is strictly smaller; the displaced car resumes its own search.
 \item \textbf{Expand only after exhausting a tour.} If a car completes its tour without remaining parked, double its priority and search the resulting component. A car starting at priority zero first tests its origin and, if necessary, moves to the smallest positive tree-edge scale.
 \item \textbf{Commit the real request.} When the virtual process consumes a previously unused space, assign the new real request to that garage, using the swapping argument above.
\end{enumerate}

The priority rule protects cars that are still searching at a small scale. If a car exhausts its searches at or below a threshold and advances beyond it, its entire component is already full of cars with priorities at most that threshold. This gives an exact count of how many cars can finish below each threshold. Any feasible offline assignment faces the same capacity restriction, so these counts allow us to charge virtual searches to the offline optimum. The comparison is between counts of cars, rather than a separate comparison for each individual car.

The remaining task is to choose the scales well. The analysis uses one quantity to bound the cost of a tour at a single scale and another to bound costs accumulated across scales. Minimizing their sum improves the guarantee. The key final argument has two cases: either the two quantities already differ enough to give a saving, or most edge lengths lie within a factor of two, where a more careful choice of phase gives the saving instead.

\section{Algorithm Description}
\label{sec:description}

If $m=1$, assign every request to the only garage. For the remaining description, assume $m\ge2$ and set $n=m-1$. Garages at the same location can be merged by adding their capacities. All choices among ties, all spanning-tree choices, and all traversal orders are fixed deterministically. The notation $\ln$ denotes the natural logarithm, and $\log_2$ denotes the base-two logarithm.

This section specifies the preprocessing and the online decisions. Their correctness and competitive guarantee are proved in Section~\ref{sec:analysis}.

\subsection{The spanning tree and the search scales}
\label{sec:scales}

Construct a minimum spanning tree $T$ of the complete graph on $G$, \emph{before any rounding}. Its edge set is $E(T)$ and it has $n$ edges. An edge $e=\{u,v\}$ has original length $w_e=d(u,v)>0$. For a phase $\theta\in[0,1)$, define the shifted dyadic scales and downward rounding by
\begin{equation}
 \mathcal R_\theta=\{2^{\theta+j}:j\in\Z\},\qquad
 \down_\theta(z)=
 \begin{cases}
 0,&z=0,\\
 2^{\theta+\lfloor\log_2z-\theta\rfloor},&z>0.
 \end{cases}
 \label{eq:round}
\end{equation}
Set $q_e=\down_\theta(w_e)$. Thus $q_e\le w_e<2q_e$, and an edge exactly on a scale rounds to itself. Rounded lengths determine which edges a search may use; traversing an edge always costs its original length $w_e$.

Define
\begin{equation}
 H(\theta)=\sum_{e\in E(T)}\frac{w_e}{q_e},\qquad
 W_\theta(r)=\sum_{\substack{e\in E(T)\\q_e\le r}}w_e,\qquad
 F(\theta)=\max_{r\in\mathcal R_\theta}\frac{W_\theta(r)}{r}.
 \label{eq:objectives}
\end{equation}

The sum $W_\theta(r)$ includes all eligible tree edges, across all components. For computing $F(\theta)$, it suffices to evaluate the ratio at the distinct rounded edge lengths.

For a real number $z$, write $\fracpart(z)=z-\lfloor z\rfloor$ for its fractional part. Form the finite set of candidate phases
\[
 \Theta=\{\fracpart(\log_2w_e):e\in E(T)\}.
\]
Evaluate $H(\theta)+F(\theta)$ for each $\theta\in\Theta$ and select a minimizer, choosing the smallest phase in case of a tie. Lemma~\ref{lem:phases} shows that this finite rule also minimizes the objective over all phases in $[0,1)$.

Fix the selected phase for the rest of the algorithm, and abbreviate $H=H(\theta)$ and $F=F(\theta)$. Define the multiplier
\begin{equation}
 \kappa=1+\frac HF.
 \label{eq:kappa}
\end{equation}
For any real threshold $u\ge0$, let $\mathcal C(u)$ be the connected components of the forest containing exactly the tree edges with $q_e\le u$. Let $T_u(g)$ be the component containing garage $g$. At $u=0$, each component is a single garage.

\subsection{The virtual algorithm and actual assignments}
\label{sec:algorithm}

For each garage $g$ and positive scale $r\in\mathcal R_\theta$, fix a closed depth-first tour of $T_r(g)$ rooted at $g$. The tour traverses every component edge twice. A singleton tour consists of one test of its garage.

When request $t$ arrives, choose a nearest garage $g_t\in G$, and define
\begin{equation}
 \delta_t=d(x_t,g_t),\qquad a_t=\down_\theta(\kappa\delta_t).
 \label{eq:initial}
\end{equation}
Introduce a virtual car with fixed origin $g_t$, initial priority $a_t$, and an itinerary rooted at $g_t$. Conceptually move it from $x_t$ to $g_t$, at cost $\delta_t$. If $a_t>0$, begin the scale-$a_t$ tour. If $a_t=0$, first test $g_t$ at priority zero; if that test is exhausted, begin at the smallest rounded tree-edge length. Completion of a positive-scale tour returns the car to its origin, doubles its priority, and starts the next tour.

At every garage test, use the following rules.
\begin{enumerate}[label=\arabic*.]
 \item If there is an unused virtual space, park the active car and finish processing the arrival.
 \item Otherwise, let $b$ be the largest priority of a resident virtual car. If the active car's priority is strictly less than $b$, replace a deterministically chosen resident of priority $b$. The incoming car parks and records its current tour position. The displaced car becomes active and resumes its own itinerary immediately after its saved parking test.
 \item If neither rule applies, the active car continues its tour.
\end{enumerate}
Each car's priority never decreases. A resumed tour continues from its saved position; it does not restart. When an arrival finishes, let $y_t$ be the garage whose virtual occupancy increased. The \emph{actual} algorithm assigns the new request $x_t$ irrevocably to $y_t$. The actual destination need not be the new virtual car's parking location.

\section{Algorithm Analysis}
\label{sec:analysis}

We first state the exact guarantee, then justify the preprocessing and the virtual implementation. Threshold counts and bounds on travel reduce the analysis to selecting the phase of the search scales.

\subsection{Constants and the main guarantee}
\label{sec:constants}

Define the absolute constants
\begin{equation}
 A=\frac{2}{\ln2},\qquad K=\frac{3}{\ln(25/8)},\qquad
 \eta=\frac{A-K}{9-2K},\qquad C=A-\eta,\qquad \rho=2C.
 \label{eq:constants}
\end{equation}
Rounded to two decimal places, their approximate values are
\[
A\approx2.89,\qquad K\approx2.63,\qquad \eta\approx0.07,\qquad
C\approx2.82,\qquad \rho\approx5.64.
\]

\begin{theorem}\label{thm:main}
For $m\ge2$ distinct garages, the deterministic algorithm in Section~\ref{sec:description} satisfies
\[
 \ALG\le\bigl[\rho(m-1)+1\bigr]\OPT.
\]
The guarantee is independent of the capacities and the number of requests. For $m=1$, the unique feasible destination gives an optimal assignment.
\end{theorem}

\subsection{Properties of the preprocessing}
\label{sec:preprocessing-analysis}

The sum $W_\theta(r)$ includes eligible edges across all components of the tree, not only the component containing a particular garage. The maximum defining $F$ is attained at some rounded edge length: below the smallest rounded length its numerator is zero, and after any occupied scale the numerator is unchanged until the next occupied scale, while the denominator increases. Also,
\begin{equation}
 0<F(\theta)\le H(\theta),
 \label{eq:FleH}
\end{equation}
since for every scale $r$,
$W_\theta(r)/r\le\sum_{q_e\le r}w_e/q_e\le H(\theta)$.
In particular, the multiplier $\kappa=1+H/F$ is well-defined and satisfies $\kappa\ge2$.

\begin{lemma}[Finite phase selection]\label{lem:phases}
A minimizing phase can be found by evaluating $H+F$ at the at most $n$ distinct phases
\[
 \theta_e=\fracpart(\log_2w_e),\qquad e\in E(T).
\]
\end{lemma}
\begin{proof}
The listed phases are exactly those at which an original edge length is a scale. Within an interval between consecutive such phases, increasing the phase by $\varepsilon$ multiplies every scale and every rounded length by $2^\varepsilon$, without changing their scale indices. Thus $H$ is multiplied by $2^{-\varepsilon}$. Matching a scale $r$ to $2^\varepsilon r$ leaves the set of included edges unchanged, so every normalized prefix sum, and therefore $F$, is also multiplied by $2^{-\varepsilon}$. Hence $H+F$ strictly decreases throughout the interval.

At an edge phase, the boundary convention retains the value reached from the left. Immediately afterward, every edge aligned with a boundary drops to the preceding scale: its contribution to $H$ jumps from $1$ to $2$ in the one-sided limit. These drops can only add edges to the limiting prefix sums, so $F$ cannot jump downward. Therefore an interval's smallest value is attained at its right-hand edge phase. Since phases $0$ and $1$ describe the same grid, the argument also applies across the endpoint of $[0,1)$. Tied edge phases are handled simultaneously.
\end{proof}

For two garages $g,h$, let $P_T(g,h)$ denote the unique tree path between them. Define their rounded bottleneck distance by
\begin{equation}
 \beta(g,h)=
 \begin{cases}
 \max_{e\in P_T(g,h)}q_e,&g\ne h,\\
 0,&g=h.
 \end{cases}
 \label{eq:beta}
\end{equation}
Thus $\beta(g,h)\le u$ exactly when $g$ and $h$ belong to the same component of $\mathcal C(u)$. The minimum spanning tree property gives
\begin{equation}
 \beta(g,h)\le\max_{e\in P_T(g,h)}w_e\le d(g,h)\qquad(g\ne h).
 \label{eq:mst}
\end{equation}
Indeed, if a path edge were longer than $d(g,h)$, adding the direct edge $\{g,h\}$ and removing that path edge would produce a lighter spanning tree. This bounds the largest edge on the path, not its total length.

\subsection{Termination, feasibility, and the swapping bound}
\label{sec:implementation-analysis}

Define $V$ to be the total length of all tree-edge movements by virtual cars over the entire input, and let $D=\sum_{t=1}^N\delta_t$.

\begin{lemma}[Termination, feasibility, and cost]\label{lem:implementation}
The process terminates on every arrival, actual assignments respect capacities, and
\begin{equation}
 \ALG\le D+V,\qquad D\le\OPT.
 \label{eq:implementation}
\end{equation}
\end{lemma}
\begin{proof}
A full virtual garage stays full. Its maximum resident priority never increases, because replacements remove a maximum-priority resident and insert a strictly smaller-priority car. If a car is displaced from a garage at priority $r$, that garage thereafter remains full with all resident priorities at most $r$. The displaced car's priority never drops below $r$, so it can never park there again. Each virtual car is therefore displaced at most $m$ times.

Only finitely many cars are present during an arrival, so only finitely many displacements can occur. If the active car keeps searching, its priority eventually reaches a scale whose component contains all garages. A free space exists because fewer than $C_{\mathrm{tot}}$ virtual cars were parked before this arrival. A full tour at that scale encounters a free space and ends the arrival. Each intervening tour is finite, and any positive starting scale reaches that scale after finitely many doublings.

Replacements preserve all occupancies; the last parking increases occupancy at exactly one garage, $y_t$. Starting from empty configurations, actual and virtual occupancies therefore agree after each arrival. In particular, $y_t$ has actual unused capacity when request $t$ is assigned there. Earlier actual assignments are never changed.

Let $V_t$ be the total tree-edge movement triggered while processing arrival $t$. The initial step from $x_t$ to $g_t$, followed by the displacement chain's movements, forms a walk from $x_t$ to $y_t$. At a displacement, the outgoing car resumes at the same garage where the incoming car stopped. The triangle inequality gives $d(x_t,y_t)\le\delta_t+V_t$. Summing and using $V=\sum_tV_t$ proves the first inequality. Every feasible destination for $x_t$ is at distance at least $\delta_t$, so summing these lower bounds over an optimal assignment proves $D\le\OPT$.
\end{proof}

\subsection{Exact counts at each priority threshold}
\label{sec:threshold}

Let $B_t$ be the final priority of virtual car $t$ after all $N$ arrivals. Both $a_t$ and $B_t$ belong to $\{0\}\cup\mathcal R_\theta$, and $B_t\ge a_t$. For a component $Q\in\mathcal C(u)$, define
\[
 c(Q)=\sum_{v\in Q}c(v),\qquad
 N_Q(u)=\bigl|\{t:g_t\in Q,\ a_t\le u\}\bigr|.
\]
The integer $c(Q)$ is its total capacity. The integer $N_Q(u)$ counts its \emph{eligible} requests: their origins lie in $Q$ and their initial priorities are at most $u$.

\begin{lemma}[Exact threshold counts]\label{lem:counts}
For every real $u\ge0$,
\begin{equation}
 \bigl|\{t:B_t\le u\}\bigr|
 =\sum_{Q\in\mathcal C(u)}\min\{N_Q(u),c(Q)\}.
 \label{eq:counts}
\end{equation}
\end{lemma}
\begin{proof}
A car with priority at most $u$ has initial priority at most $u$ and has traversed only edges with rounded length at most $u$. It must therefore lie in its origin's component in $\mathcal C(u)$.

Suppose an eligible car raises its priority above $u$. Before doing so, it must exhaust a tour of its entire origin component $Q$. For $u>0$ this is the tour at the largest dyadic scale at most $u$, unless the only relevant component is a singleton before the first positive search level. That singleton is exhausted by a failed priority-zero test. For $u=0$, the same priority-zero test suffices.

At each garage visited on the exhausted tour, one of two events lets the car continue. Either its test fails, in which case the garage is full and its maximum priority is no larger than the car's; or the car parks and is later displaced as a maximum-priority resident, leaving only residents of no larger priority. Since the tour's priority is at most $u$, every garage in $Q$ is now full of cars with priorities at most $u$. This condition persists because maxima at full garages cannot increase. Call $Q$ \emph{sealed at threshold $u$}.

If $N_Q(u)\le c(Q)$, no eligible car from $Q$ can advance above $u$: the first such advance would leave $c(Q)$ eligible cars parked in $Q$ in addition to the advancing car, contradicting the total count $N_Q(u)$. Thus all $N_Q(u)$ eligible cars finish at priority at most $u$.

If $N_Q(u)>c(Q)$, at least one eligible car must finish above $u$, because at most $c(Q)$ cars of priority at most $u$ can be parked in $Q$. Its advance seals $Q$. Exactly $c(Q)$ cars then finish in $Q$ with priorities at most $u$, and all of them originate in $Q$. Thus each component contributes exactly the minimum in~\eqref{eq:counts}. Summing over components proves the claim.
\end{proof}

For any feasible assignment sending request $t$ to garage $h_t$, define its \emph{extended bottleneck cost} for that request by
\begin{equation}
 U_t=\max\{a_t,\beta(g_t,h_t)\}.
 \label{eq:extended}
\end{equation}
This is an auxiliary score, distinct from the actual metric cost $d(x_t,h_t)$. The condition $U_t\le u$ requires both eligibility ($a_t\le u$) and assignment within the origin's component of $\mathcal C(u)$.

\begin{lemma}[Threshold domination]\label{lem:domination}
For any feasible assignment with extended costs $U_1,\ldots,U_N$ and every $u\ge0$,
\begin{equation}
 \bigl|\{t:B_t>u\}\bigr|\le\bigl|\{t:U_t>u\}\bigr|.
 \label{eq:domination}
\end{equation}
\end{lemma}
\begin{proof}
In a component $Q$, at most $N_Q(u)$ requests are eligible and at most $c(Q)$ can be assigned there. Any feasible assignment therefore has at most $\min\{N_Q(u),c(Q)\}$ pairs of extended cost at most $u$ originating in $Q$. Summing and applying Lemma~\ref{lem:counts} gives
\[
 |\{t:U_t\le u\}|\le\sum_Q\min\{N_Q(u),c(Q)\}
 =|\{t:B_t\le u\}|.
\]
Subtract both sides from $N$.
\end{proof}

The lemma compares \emph{counts}, not the priorities of corresponding individual cars. It does not assert $B_t\le U_t$ for each $t$.

\subsection{Bounding the total virtual travel}
\label{sec:travel}

Fix an optimal assignment in the original metric, and let $h_t^*$ be its destination for request $t$. Define
\begin{equation}
 d_t=d(x_t,h_t^*),\qquad
 b_t=\down_\theta(\delta_t+d_t),\qquad
 M_t=\max\{a_t,b_t\}.
 \label{eq:comparison}
\end{equation}
Then $\delta_t\le d_t$ and $\sum_td_t=\OPT$. By~\eqref{eq:mst} and the triangle inequality,
\[
 \beta(g_t,h_t^*)\le d(g_t,h_t^*)\le\delta_t+d_t.
\]
The first expression is zero or an element of $\mathcal R_\theta$. Since $b_t$ is the largest scale no greater than $\delta_t+d_t$, we obtain $\beta(g_t,h_t^*)\le b_t$. Consequently the optimal assignment's extended cost
$U_t=\max\{a_t,\beta(g_t,h_t^*)\}$ is at most $M_t$.

For a grid scale $r$, all of $B_t,U_t,M_t$ are zero or grid values. There is no grid value strictly between $r/2$ and $r$. Applying Lemma~\ref{lem:domination} at $u=r/2$ therefore gives
\begin{equation}
 |\{t:B_t\ge r\}|=|\{t:B_t>r/2\}|
 \le|\{t:U_t>r/2\}|\le|\{t:M_t\ge r\}|.
 \label{eq:gridcounts}
\end{equation}

\begin{lemma}[Charging tours by scale]\label{lem:tourcharge}
The total tree-edge movement satisfies
\begin{equation}
 V\le\sum_{t=1}^N 2\sum_{\substack{r\in\mathcal R_\theta\\a_t\le r\le M_t}}W_\theta(r).
 \label{eq:tourcharge}
\end{equation}
\end{lemma}
\begin{proof}
A car performs at most one tour at any positive scale, counting all resumptions as parts of that same tour. A scale-$r$ tour costs at most $2W_\theta(r)$ because its component edges are a subset of the edges in $W_\theta(r)$. The number of cars that perform such a tour is at most
\[
 |\{t:a_t\le r\le B_t\}|=|\{t:B_t\ge r\}|-|\{t:a_t>r\}|,
\]
where the equality uses $B_t\ge a_t$. By~\eqref{eq:gridcounts}, this is at most
\[
 |\{t:M_t\ge r\}|-|\{t:a_t>r\}|=|\{t:a_t\le r\le M_t\}|,
\]
using $M_t\ge a_t$. Multiply by $2W_\theta(r)$, sum over scales, and interchange the sums. Although the grid extends toward zero, $W_\theta(r)=0$ below the smallest rounded edge length. All nonzero terms in the calculation form a finite sum.
\end{proof}

\begin{lemma}[Accumulating scales]\label{lem:accumulation}
For any positive grid level $B\in\mathcal R_\theta$,
\begin{equation}
 \sum_{\substack{r\in\mathcal R_\theta\\r\le B}}W_\theta(r)\le HB.
 \label{eq:accumulation}
\end{equation}
\end{lemma}
\begin{proof}
Here $B$ is an arbitrary cutoff level; it is not the final priority $B_t$ of a specified car. Fix an edge with $q_e\le B$ and write $B/q_e=2^j$, where $j\ge0$ is an integer. It contributes $w_e$ at exactly the levels $q_e,2q_e,\ldots,2^jq_e=B$, hence $j+1$ times. Since $j+1\le2^j$,
\begin{align*}
 \sum_{r\le B}W_\theta(r)
 &=\sum_{e:q_e\le B}w_e\left(1+\log_2\frac{B}{q_e}\right)\\
 &\le B\sum_{e:q_e\le B}\frac{w_e}{q_e}
 \le B\sum_{e\in E(T)}\frac{w_e}{q_e}=BH.
\end{align*}
The sum over levels is over $\mathcal R_\theta$. Edges with $q_e>B$ do not occur on the left; including them on the right only enlarges the bound.
\end{proof}

\begin{theorem}[Cost for a fixed phase]\label{thm:fixedphase}
For any fixed phase, using $\kappa=1+H/F$ gives
\begin{equation}
 \ALG\le\bigl[2(H+F)+1\bigr]\OPT.
 \label{eq:fixedphase}
\end{equation}
\end{theorem}
\begin{proof}
For each optimal assignment pair, define its charge in Lemma~\ref{lem:tourcharge} by
\[
 Q_t=2\sum_{\substack{r\in\mathcal R_\theta\\a_t\le r\le M_t}}W_\theta(r).
\]
We bound $Q_t$ in terms of $d_t$. This is an accounting charge, not a bound on the movement of that individual virtual car.

If $b_t\le a_t$, then $M_t=a_t$. If $a_t=0$, the sum has no positive scale and $Q_t=0$. Otherwise there is only one scale to charge, and
\[
 Q_t=2W_\theta(a_t)\le2Fa_t\le2\kappa F\delta_t\le2\kappa Fd_t.
\]

If $b_t>a_t$, put $B=b_t>0$. When $a_t>0$, adjacent grid levels give $B\ge2a_t$, while downward rounding gives $a_t>\kappa\delta_t/2$. Thus $B>\kappa\delta_t$. If $a_t=0$, then $\delta_t=0$, so again $\delta_t<B/\kappa$. Since $B\le\delta_t+d_t$,
\[
 d_t\ge B-\delta_t>\left(1-\frac1\kappa\right)B.
\]
Lemma~\ref{lem:accumulation}, allowing all lower levels in the upper bound, yields
\[
 Q_t\le2\sum_{r\le B}W_\theta(r)\le2HB
 \le\frac{2\kappa H}{\kappa-1}d_t.
\]
The definition of $\kappa$ balances the two coefficients:
\[
 2\kappa F=\frac{2\kappa H}{\kappa-1}=2(H+F).
\]
Hence $V\le\sum_tQ_t\le2(H+F)\sum_td_t=2(H+F)\OPT$. Lemma~\ref{lem:implementation} and $D\le\OPT$ finish the proof.
\end{proof}

\subsection{A phase bound for lengths within a factor of two}
\label{sec:oneoctave}

We now bound the best value of $H+F$. The next lemma concerns a collection of $p$ positive lengths, which may later be a subset of the tree edges. The quantities $H$, $W_\theta$, and $F$ are computed from that collection alone by the formulas in~\eqref{eq:objectives}. In particular, $p$ counts lengths, not requests or capacity units.

\begin{lemma}[One-octave bound]\label{lem:octave}
Let $w_1,\ldots,w_p$ be positive lengths in $[a,2a]$, where $a>0$ and $p\ge1$. Then
\begin{equation}
 \min_{\theta\in[0,1)}\bigl(H(\theta)+F(\theta)\bigr)\le Kp,
 \qquad K=\frac3{\ln(25/8)}.
 \label{eq:octave}
\end{equation}
\end{lemma}
\begin{proof}
Rescaling every length by $1/a$ merely shifts the available phases and rescales the associated grid. The ratios in $H$ and $F$ are unchanged under the corresponding rescaling, so assume $1\le w_i\le2$. Define
\[
 W_{\mathrm{tot}}=\sum_{i=1}^p w_i,\qquad
 s(t)=\frac{\sum_{i:w_i<t}w_i}{W_{\mathrm{tot}}}\quad(1<t<2).
\]
Thus $W_{\mathrm{tot}}$ is the total unrounded length, and $s(t)$ is the fraction of that total contributed by lengths below the cutoff $t$. The variable $t$ is a real cutoff in this proof, not a request index.

Choose the phase whose grid contains $t$. Every length below $t$ rounds to $t/2$, and every length at least $t$ rounds to $t$. Writing $s=s(t)$, we obtain
\[
 H=\frac{sW_{\mathrm{tot}}}{t/2}+\frac{(1-s)W_{\mathrm{tot}}}{t}
 =\frac{W_{\mathrm{tot}}}{t}(1+s).
\]
The only potentially maximizing prefix scales for $F$ are $t/2$ and $t$. Their values are $2sW_{\mathrm{tot}}/t$ and $W_{\mathrm{tot}}/t$, respectively. Therefore
\begin{equation}
 H+F=\frac{W_{\mathrm{tot}}}{t}\,g(s(t)),\qquad
 g(s)=\begin{cases}
 2+s,&0\le s\le1/2,\\
 1+3s,&1/2\le s\le1.
 \end{cases}
 \label{eq:gs}
\end{equation}
The strict condition $w_i<t$ is intentional: at equality the edge rounds to $t$, so it belongs to the upper group.

Let $M=\min_\theta(H+F)$ for these $p$ lengths, and set $z=M/W_{\mathrm{tot}}$. Both are positive. Since $g(s)\le4$, taking $t$ toward $2$ in~\eqref{eq:gs} shows $z\le2$. Every cutoff has objective at least $M$, hence $g(s(t))\ge zt$. Inverting the increasing function $g$ on its range gives
\begin{equation}
 s(t)\ge\varphi(zt),\qquad
 \varphi(y)=\begin{cases}
 0,&0\le y\le2,\\
 y-2,&2\le y\le5/2,\\
 (y-1)/3,&5/2\le y\le4.
 \end{cases}
 \label{eq:phi}
\end{equation}
No arguments above $4$ occur because $z\le2$ and $t\le2$. Thus a large minimum objective forces a lower bound on the accumulated length below every cutoff.

We next convert accumulated length into a count of lengths. For every $i$,
\[
 \int_{w_i}^2\frac{dt}{t^2}=\frac1{w_i}-\frac12.
\]
The contribution of length $w_i$ to $s(t)$ is $w_i/W_{\mathrm{tot}}$ when $t>w_i$ and zero otherwise. Integrating these contributions separately gives the exact identity
\begin{align}
 \int_1^2\frac{s(t)}{t^2}\,dt
 &=\frac1{W_{\mathrm{tot}}}\sum_{i=1}^p
     w_i\int_{w_i}^2\frac{dt}{t^2}\notag\\
 &=\frac1{W_{\mathrm{tot}}}\sum_{i=1}^p
      \left(1-\frac{w_i}{2}\right)
 =\frac{p}{W_{\mathrm{tot}}}-\frac12.
 \label{eq:countintegral}
\end{align}
The value at a single cutoff does not affect an integral, so this also covers lengths equal to $1$ or $2$ and repeated lengths.

Define the auxiliary function
\begin{equation}
 \mathcal D(z)=\frac12+\int_1^2\frac{\varphi(zt)}{t^2}\,dt.
 \label{eq:Ddefinition}
\end{equation}
This function is unrelated to the earlier total distance $D=\sum_t\delta_t$. Equations~\eqref{eq:phi} and~\eqref{eq:countintegral} imply $p/W_{\mathrm{tot}}\ge\mathcal D(z)$. The integration changes form when $zt$ crosses $2$ or $5/2$. Since $1\le t\le2$ and $z\le2$, the resulting expressions are
\begin{equation}
 \mathcal D(z)=\begin{cases}
 \dfrac12,&0<z\le1,\\[1mm]
 \dfrac12+\displaystyle\int_{2/z}^2\frac{zt-2}{t^2}\,dt
 =z\ln z+\dfrac32-z,&1\le z\le5/4,\\[2mm]
 \dfrac12+\displaystyle\int_{2/z}^{5/(2z)}\frac{zt-2}{t^2}\,dt
 +\displaystyle\int_{5/(2z)}^2\frac{zt-1}{3t^2}\,dt,&5/4\le z\le2.
 \end{cases}
 \label{eq:Dintegrals}
\end{equation}
In the last interval the expression simplifies to
\[
 \mathcal D(z)=z\ln(5/4)+\frac z3\ln(4z/5)+\frac{2-z}{3}.
\]
On the three intervals above, the derivative of $\mathcal D(z)/z$ is, respectively,
\[
 -\frac1{2z^2},\qquad\frac{2z-3}{2z^2},\qquad\frac{z-2}{3z^2}.
\]
Each derivative is nonpositive in its interval, and the expressions agree at $z=1$ and $z=5/4$. Hence $\mathcal D(z)/z$ is nonincreasing on $(0,2]$. Using $M=zW_{\mathrm{tot}}$ gives
\begin{align*}
 \frac pM
 &=\frac{p/W_{\mathrm{tot}}}{z}
 \ge\frac{\mathcal D(z)}z
 \ge\frac{\mathcal D(2)}2\\
 &=\ln(5/4)+\frac13\ln(8/5)
 =\frac13\ln(25/8).
\end{align*}
Rearranging proves $M\le3p/\ln(25/8)=Kp$.
\end{proof}

\subsection{The joint phase bound and completion of the proof}
\label{sec:joint}

\begin{lemma}[Joint phase bound]\label{lem:joint}
For any $n\ge1$ positive edge lengths, the objectives in~\eqref{eq:objectives} satisfy
\begin{equation}
 \min_{\theta\in[0,1)}\bigl(H(\theta)+F(\theta)\bigr)\le Cn,
 \label{eq:joint}
\end{equation}
where $C$ is defined in~\eqref{eq:constants}.
\end{lemma}
\begin{proof}
For a fixed edge and a uniformly chosen phase,
\[
 \frac{w_e}{q_e}=2^{\fracpart(\log_2w_e-\theta)},\qquad
 \int_0^1\frac{w_e}{q_e}\,d\theta=\int_0^1 2^u\,du=\frac1{\ln2}.
\]
The integration variable $u$ in this display is a fractional logarithmic position. Summing over edges shows that some phase $\theta_0$ has $H(\theta_0)\le n/\ln2$. At this phase, write $\Delta=H(\theta_0)-F(\theta_0)\ge0$.

If $\Delta\ge\eta n$, then
\[
 H(\theta_0)+F(\theta_0)=2H(\theta_0)-\Delta
 \le\left(\frac2{\ln2}-\eta\right)n=Cn.
\]
Here the gap between $H$ and $F$ directly supplies the improvement.

Suppose instead that $\Delta<\eta n$. Choose a scale $r_0$ attaining $F(\theta_0)=W_{\theta_0}(r_0)/r_0$. In the remainder of this paragraph, each $q_e$ is rounded using $\theta_0$. Let $I=\{e:q_e=r_0\}$ and let $n_{\mathrm{out}}$ count the edges outside $I$. Edges in $I$ contribute equally to $H$ and this maximizing prefix sum. If $q_e<r_0$, then $q_e\le r_0/2$, so its contribution to their difference is
\[
 \frac{w_e}{q_e}-\frac{w_e}{r_0}
 \ge\frac12\frac{w_e}{q_e}\ge\frac12.
\]
If $q_e>r_0$, it contributes zero to the prefix sum and at least $1$ to $H$. Every edge outside $I$ therefore contributes at least $1/2$ to $\Delta$, giving
\begin{equation}
 n_{\mathrm{out}}\le2\Delta<2\eta n.
 \label{eq:outside}
\end{equation}
Thus a small gap forces most edges into a single bin.

The original lengths of the $n-n_{\mathrm{out}}$ edges in $I$ lie in $[r_0,2r_0)$. Apply Lemma~\ref{lem:octave} to this fixed subset and choose a phase $\theta_1$ for which its objective is at most $K(n-n_{\mathrm{out}})$. To combine it with the remaining edges, for any subset $J$ define $H_J$, $W_{J,\theta}$, and $F_J$ by restricting the sums in~\eqref{eq:objectives} to $J$. At any common phase and for disjoint subsets $J,L$,
\[
 H_{J\cup L}=H_J+H_L,\qquad
 F_{J\cup L}=\max_r\frac{W_{J,\theta}(r)+W_{L,\theta}(r)}r
 \le F_J+F_L.
\]
For a single edge, both its $H$ and its $F$ equal $w_e/q_e<2$, at every phase. Its separate objective is therefore less than $4$. Subadditivity and~\eqref{eq:outside} give
\begin{align*}
 H(\theta_1)+F(\theta_1)
 &\le K(n-n_{\mathrm{out}})+4n_{\mathrm{out}}\\
 &=Kn+(4-K)n_{\mathrm{out}}\\
 &\le\bigl[K+2(4-K)\eta\bigr]n=Cn.
\end{align*}
The last equality follows from $\eta=(A-K)/(9-2K)$, which balances the two cases:
$K+2(4-K)\eta=A-\eta=C$. Thus either $\theta_0$ or $\theta_1$ achieves the claimed bound. Minimizing $H+F$ performs at least as well as either witness phase.
\end{proof}

\begin{proof}[Proof of Theorem~\ref{thm:main}]
The minimum spanning tree has $n=m-1$ edges. By Lemma~\ref{lem:joint}, the selected phase has $H+F\le C(m-1)$. Theorem~\ref{thm:fixedphase} consequently gives
\[
 \ALG\le[2(H+F)+1]\OPT
 \le[2C(m-1)+1]\OPT
 =[\rho(m-1)+1]\OPT.
\]
The phase, tree, and tie-breaking rules are selected deterministically before requests arrive, and each online decision uses only the current request and maintained state. Averaging over phases is used solely to prove that a good preprocessing choice exists. For $m=1$, all requests have the same unique feasible garage and the algorithm is optimal.
\end{proof}

\section{AI methodology}
\label{sec:ai-disclosure}

\subsection*{Activities}

The following checklist describes the use of AI in this work. The first four items are at the level of a human co-author, while the last two items are at the level of any substantial contribution.

\begin{itemize}
    \item Asking the research question - No
    \item Coming up with the approach - Yes
    \item Proof development - Yes
    \item Writing and exposition - Yes
    \item Checking for bugs - Yes
    \item Other supporting tasks - Yes
\end{itemize}

\subsection*{Details}
After we suggested to improve upon the virtual routing found in  \cite{ampu26}, OpenAI's ChatGPT-6 Astra assisted with developing the algorithmic ideas and proof arguments. In particular, while we believed it possible to adjust the edge-rounding procedure, the optimization presented in this work would not be possible without OpenAI's ChatGPT. We manually checked all proofs and believe them to be correct, however the idea to have a \textit{second} objective function $F$, over which the algorithm jointly minimizes in tandem with the global distortion $H$ was a novel idea produced by ChatGPT. Moreover the original draft was entirely created by ChatGPT. We made several organizational adjustments, aided by further prompting ChatGPT, to get a more intuitive draft on which we made manual edits and revisions. Finally, ChatGPT directed us to some additional references, all of which we have checked for relevance.

\begingroup
\small
\bibliographystyle{unsrt}
\bibliography{references}
\endgroup
\end{document}